# Autonomics: In Search of a Foundation for Next Generation Autonomous Systems

David Harel[1], Assaf Marron[1] and Joseph Sifakis[2]

[1] Weizmann Institute of Science, Rehovot, Israel
[2] Univ. Grenoble Alpes, Verimag laboratory, St Martin d'Heres, France

david.harel@weizmann.ac.il   assaf.marron@weizmann.ac.il   joseph.sifakis@imag.fr

**Abstract**

The potential benefits of autonomous systems have been driving intensive development of such systems, and of supporting tools and methodologies. However, there are still major issues to be dealt with before such development becomes commonplace engineering practice, with accepted and trustworthy deliverables. We argue that a solid, evolving, publicly available, community-controlled foundation for developing next generation autonomous systems is a must. We discuss what is needed for such a foundation, identify a central aspect thereof, namely, decision-making, and focus on three main challenges: (i) how to specify autonomous system behavior and the associated decisions in the face of unpredictability of future events and conditions and the inadequacy of current languages for describing these; (ii) how to carry out faithful simulation and analysis of system behavior with respect to rich environments that include humans, physical artifacts, and other systems,; and (iii) how to engineer systems that combine executable model-driven techniques and data-driven machine learning techniques. We argue that *autonomics*, i.e., the study of unique challenges presented by next generation autonomous systems, and research towards resolving them, can introduce substantial contributions and innovations in system engineering and computer science.

## 1 Introduction

Many autonomous systems are already able to replace humans in carrying out a variety of functions. This trend will continue in the years to come, resulting in autonomous systems becoming a central and crucial part of human society in numerous areas. These will include vehicles of all kinds, medical and industrial robots and assistants, agricultural and manufacturing facilities, and distributed management for traffic, urban security, electric grids, etc.

Many organizations are already striving to develop trustworthy, cost-effective autonomous systems. To make this possible, powerful tools and methods are being researched. However, due to the expected increase in both autonomy and criticality of such systems, we believe that the required trustworthiness is not achievable with current practices [25]. Consider, for example, even a very modest autonomous system, a valet-parking robot – obviously a far cry from a full autonomous vehicle. Customers and regulators would be fully justified in asking whether the robot will be able to discover a child forgotten in the car, or notice that a house pet is sleeping underneath a parked vehicle. And if the robot is indeed able to notice these, what will it do as a result? How will the robot react if a human attempts to stop it by pursuing it and yelling, or by blocking its way? Developers of such systems must not only solve such challenges and have good answers when asked about such scenarios; they must be able to readily convince wide audiences that their systems can be trusted under a wide range of unexpected, not-yet-articulated conditions.

We argue that such development challenges cannot be dealt with by enhancing the system's safety features and merely adding, say, sensors, dedicated safety or decision making components, and test cases. The new levels complexity and criticality present fundamental new challenges for researchers, engineers, and regulators. The challenge can be summarized as follows:

> *Next-generation autonomous systems will be expected to operate under conditions which will be unpredictable at the time of their development. This unpredictability is due to the limited controllability over the system's environment, the emergence of new objects, properties and events in the world, and the exponential explosion of the number of composite configurations of such conditions and events. How can engineers assure customers and regulators that the system will function correctly and safely, before many complex conditions are even conceived? Matters are further complicated by the facts that many conditions which can be conceived still cannot be described with current languages, and when they can be described, it is not clear what instructions one should give the systems for handling then.*

We believe that to close this gap between the challenges in developing trustworthy next-generation autonomous systems and the existing research and technologies (in fields like programming languages, formal methods, system engineering, automation, control, artificial intelligence and machine learning, smart-agent and multi-agent systems) our community needs to construct a common engineering foundation for developing such systems. This *autonomics* foundation will study the unique challenges presented by next generation autonomous systems, and provide engineering principles, methods, tools and examples, as well as means for selecting among multiple design alternatives. Such a foundation will dramatically accelerate the development and acceptance of high quality, certifiable autonomous systems, built for the benefit of human society.

In Section 2, we describe the principles we believe will characterize next-generation autonomous systems, and how they differ from those of present-day systems. The section will offer a basis and common terminology for the rest of the paper. In Section 3, we zoom in on one particular aspect of the broad problem – the development of decision-making processes, and identify three key gaps: (i) specifying autonomous system behavior and the decisions to be made therein, when working under unpredictable evolving environments, and validating that the resulting decisions will indeed meet what the users would have wanted the system to do; (ii) carrying out adequate simulations and analysis that would serve to validate a system's actual decisions as observed under complex situations; and (iii) designing means for hybrid decision-making that combines data-based techniques (based on AI and machine learning) with algorithmic, executable model-based ones. In Section 4 we provide a brief review of related work, in an attempt to recognize the vast amount of work that has already done, and at the same time support our argument about the remaining challenges.

## 2  Characterizing Next Generation Autonomous Systems

### 2.1  Basic Terminology

We first define our basic terms and concepts, in order to set a common ground for the ensuing discourse (see also [24, 26]). We illustrate these with an example of an autonomous vehicle for factory-floor and plant yard deliveries (termed here FFAV).

- **Systems** are the main subject of study in this paper, and are the artefacts that development teams are out to develop. A system works within an external environment, and it consists of two types of components, agents and objects, which, as we shall see, operate within a common internal system environment. The coordinated collective behavior of the system's agents and objects is designed to meet some global, system-wide, goals.

- **Objects** are those components whose behavior is not affected during system development. For example, objects of the FFAV system may include OEM-made components, such as a motor, a set of cameras, or a steering mechanism controlling the angle of the front wheels whose input is just the desired angle or a turning force (as opposed to making the decision to what angle to turn the wheels). A system often also interacts with objects that are not part of it, but are part of the environment. Such are, in the case of the FFAV, machines on the factory floor (which may serve as mere obstacles, or as recipients of deliveries), packages to be delivered, and a website from which software updates may be automatically downloaded. Objects have *states*, which can be changed by agents or by other objects, or can change 'spontaneously', for their own internal reasons.

- **Agents** are the main elements of an autonomous system. They are those components that are designed (programmed, built) as part of the system's development process.[1] Agents have *agency*: they are proactive and pursue specific goals which may change dynamically. Agents can monitor objects from the internal and external environments and can change their states, They can also coordinate their own actions with other agents. Thus, the FFAV system may either have a single agent for all its functions, or separate agents for different tasks, such as work scheduling, route planning, travel control, gripping and carrying, and delivery control. Agents can themselves be autonomous systems, and can, in turn, hierarchically contain other agents and systems.

- **The internal environment** is the lower-level physical and virtual infrastructure used by the system's agents and objects. It may include the computer/processor/memory, batteries and other power sources, the operating system, communication hardware and software, data-base management software, etc.

---

[1] We ignore here the question of whether a given component, already developed and then incorporated 'as is' into the system, should be considered as an agent or an object thereof. Similarly, we sidestep the question of whether agents that are part of systems in the external environment, like those of autonomous manufacturing machines in the context of the FFAV, should be considered agents or objects.

- **The external environment** of a system (often referred to simply as its **environment)**, is the collection of all entities with which the system might interact. It may include other systems (with their objects and agents), stand-alone objects, and any other physical or virtual entities that may affect, or be affected by, the system's behavior, via, e.g., sensors and actuators, or in other ways.
- **The ecosystem** is a general term that we use, when discussing the operation of a system in the actual world, to stand for a collection of physical and virtual entities that include the system and its external environment, and any other entities that may be relevant to such a discussion.

## 2.2 Characterizing Autonomous Behavior

We say that a system or an agent (for simplicity, we shall stick to system below) manifests autonomous behavior if it embodies the following five behavioral functions, and these are carried out with little or no intervention from humans or from other systems.

Two functions are combined to enable the system to build for itself a useful representation of the state of the external environment. **Perception** is the function needed to input stimuli, interpret their basic meaning, remove ambiguity and vagueness from complex inputs, yielding relevant information. Perception often employs, in addition to data from other systems, multi-modal inputs, such as vision, sound, heat, touch, radar, and data communication from other systems, which are obtained using mode-specific sensors and input devices. The system then amalgamates the received information. The second function is **Reflection**, which uses the information provided by the Perception function to create and constantly update an integrated run-time model representing the system's environment and its states. This model will then be used in on-going decision making.

Two other functions constitute together the system's adaptive decision process. Being adaptive means that decisions consider many possibly conflicting goals, in a way that depends on the current state of the environment. The **Goal Management** function chooses from among the set of goals the ones that are relevant to the current situation. The second function is **Planning**, which, computes a plan to achieve the set of goals produced by Goal Management, subject to state-dependent constraints. The plan is the agent's action in response to the current environment state, and may consist of, e.g., a sequence of commands to be executed by actuators.

The fifth function that characterizes autonomous behavior is **Self-Adaptation**. This is the possibility of dynamic adjustment, over time, of the system's goals and the goal management and planning processes, through learning and reasoning, based on the evolving state of the system and its environment.

## 2.3 How Next Generation Autonomous Systems Differ From Existing Systems

Next-generation autonomous systems, both those that are already beginning to emerge and definitely those of the future, differ from existing systems in the several key aspects.

1. <u>They have a large variety of possibly-conflicting system goals</u>. A typical next generation autonomous system will not be focused on a small number of well-defined goals, such as winning a game of chess, or a vehicle reaching a destination without collisions and without breaking the law. Instead, such systems will normally face a far wider and more elaborate set of goals, as humans often do. Consider, for example, the FFAV making a highly critical (and expensive) delivery, which may be at risk due to a safety issue. The difficult decision is further complicated by the business and legal considerations and risks of the company that made (and programmed) the FFAV[2].To some extent, the existence multiple, possibly conflicting, goals reflects the transition from "weak AI" to "strong AI", which contrasts specific problem-solving with devising general decision methods for complex combinations of goals. However, we believe that strong AI is not a prerequisite to trustworthy autonomous systems of the future: individual systems will be able to solve complex issues as in the above FFAV example, even without having at their disposal general strong AI solutions.

2. <u>Their environment is dramatically less predictable</u>. Autonomous systems under development already have to deal with a multitude of known environment configurations; e.g., for an autonomous vehicle, combinations of road-intersection topology, traffic volume, various driver profiles, and even the urgency of completing the trip at hand. These kinds of considerations can probably be weighed and reconciled using current or near-term technology. In contrast, next generation autonomous vehicles will face far more complex issues, which human drivers handle routinely but autonomous system will have a hard time tackling. Such hard-to-predict issues (which are also hard-to-specify and hard-to-handle when predicted) include, e.g., the whims of bicycle and motorcycle riders, weaving in and out of traffic on roads, sidewalks and crosswalk; police instructions, whether spoken or signaled; poorly marked temporary diversions; and, emergencies which have not yet been handled by first responders, such as a traffic accident, a landslide/rock-fall or flooding; or, an urgent request by a passenger to stop and step out, but where there is no safe place to do so. And, when a human is called upon to intervene and help in a difficult decision, the system must be able to handle communication issues, human inattention, or even confusion and indecision. Further, many of these conditions can happen together with each other.

   The causes of this increase in difficulty in autonomous system design include: (a) increased dependency on (hard-to-predict) physical aspects of the environment, and sensitivity to changes therein; (b) increased mobility (and an increase in sheer numbers), yielding encounters with a multitude of diverse physical locations with their own unpredictable conditions; (c) increased physical distribution of a multi-component system, creating complex combination of states at all the relevant locations; (d) increased exposure to adversarial human actions, which includes, among other things, systematic exploration of system vulnerabilities and of ways to generate undesired system behavior.

---

[2] Compare this with the difficulty of a human in choosing between two options, considering not only on the merits and risks of these options, but also personal ramifications, should others disagree with their choice.

3. <u>They require rich collaboration with humans</u>. The human-interfaces needed for next-generation autonomous systems will far exceed what is available in classical HCI, which is typically geared towards the trained user or operator *controlling* an automated task. Future interfaces will have to deal with the overall behavior of the system as seen, heard and otherwise sensed by humans, with the way the system affects human behavior and with the way humans think about system behavior (e.g., risk consideration). One thing that drives this shift is that future system missions will affect and put at risk a far wider circle of people. To illustrate these issues consider a traffic jam caused by an autonomous vehicle on a busy highway, the autonomous piloting of a plane, an autonomous crane in a busy construction site in a city center, or a medical-supply delivery robot, scurrying in a crowded hospital corridor.

   A second driver of complexity is that because a system will operate in an everyday human environment, its behavior will not only have to *be* functional, efficient and safe, but will have to *appear* to be so, and to instill in humans the *confidence* that it is indeed so. Thus, if human drivers and pedestrians will avoid passing robots or autonomous vehicles more than they avoid other pedestrians or human-driven vehicles at present, or conversely, they will allow themselves to take more liberties with such autonomous systems, then these systems will in fact be interfacing with humans in wholly new ways. And this is, of course, true not only for avoiding collisions: humans will routinely be interacting with machines on numerous everyday routines like negotiating the right-of-way through an office door, or pointing out a spill on the floor to a passing cleaning assistant.

   A third factor is collaboration. The human-computer interface is no longer just a display and keyboard---it becomes what the autonomous systems does and what it can understand. If the FFAV has to hand a delicate package to a human, and take from them another package, how does it communicate intentions (e.g., its desire to hand over one package and receive another), its questions (e.g., has the human recipient already secured a hold on the first package?), its state (e.g., that it is now holding the second package safely so that the human's hold and attention is no longer needed), etc.

   Fourth, special attention has to be given to those parts of the interface that allow a human to interrupt the operation of the autonomous system or change it abruptly. If a worker just dropped a contact lens that the FFAV cannot see, how does he/she immediately stop it? If workers gave the FFAV a wrong package, how do they call it back? If some emergency work blocks the normal, pre-programmed route of the FFAV, and a detour cannot be just discovered by trial and error exploration, how does one give the FFAV an alternate, ad-hoc instruction like "use the handicap ramp behind building C"?

## 2.4 The Issue is not *Executable Models vs. Machine Learning*

The differences between machine learning, data-driven techniques, and those based on model-driven executable specifications, are an important ingredient of this paper, and of the proposed autonomics foundation. However, we do not focus on these differences, but, rather, on the challenges in the development of next generation autonomous systems as described above. To illustrate this "difference between differences", we claim that even if next generation autonomous systems were to use only model-based and traditional procedural programming approaches, or, if all machine learning open issues, including accuracy, explainability, correctness proofs, vulnerabilities, etc., were to be resolved, many of the challenges we discuss would remain.

## 3  Why do We Need a New Foundation?

The main claim in this paper is that there are fundamental issues that need to be addressed for the development of trustworthy next-generation autonomous systems, but which have not yet been dealt with adequately by present research or by industrial experience. We call upon the research and engineering community, both in academia and in industry, to create and evolve a foundation for developing such systems. This foundation will recommend engineering practices and methods, point at tools and technologies, and offer open-source bases or examples. It will also include meta-information, such as reliable means for selecting among various system design and development alternatives.  While the autonomics foundation should touch upon all aspects of system engineering, including requirements, design, development, validation/verification/testing, certification and data collection, it should not aim to rewrite well-accepted system engineering principles. Instead it should be built (and perhaps also organized) to address the 'burning' issues and questions in engineering next generation autonomous systems.

To reinforce the argument about the depth and urgency of the required foundation, we focus in this section on one central aspect, which is at the very heart of autonomous system engineering – the decision-making. We present three partially-overlapping challenges that engineers face when developing decision-making processes, and for which satisfactory solutions are yet to come.

### 3.1  Specifying Behavior

A prerequisite for any engineering process is the ability to describe the 'hoped-for' outcome of the endeavor. In software and systems engineering this calls for specifying the system's behavior in some formal terms with agreed-upon semantics, or at the very least in very precise and technically-oriented natural-language text. This is needed in virtually all stages and (possibly agile) activities of the process: requirements, design, simulation, testing, verification, and validation. Although numerous languages have been developed for this purpose – procedural, declarative logic-based, scenario-based, and more – we argue that in next-generation autonomous systems the very specification of desired, undesired, or even actual observed behavior introduces new challenges that call for extensive research and development of languages and tools.  These challenges include the following:

1. **<u>Individual Goal Specification.</u>**  When it comes to complex autonomous systems, the specification of even a simple single goal is hard. Assume that, for the first time, an employer wants to replace a cleaning person with a robot. What kinds of specification are we after? Should it be action oriented (e.g., where and how to sweep), 'object' oriented (e.g., what kinds of dirt, dust and hair, etc., should be removed, and where from), or result oriented (e.g., what should the floors and shelves look like once the job is done)? How should one tell developers (and the system) about the need to move small objects, or unplug devices that are in the way, or about dealing with risks like breaking something?  We believe that the answer lies in comprehensive domain-specific ontologies of objects, properties, actions and relations, with textual and graphic languages (which may not be very different from current languages), for, so to speak, describing the world and its associated behaviors. A key feature of these languages should be powerful abstraction and generalization, which should alleviate the need to specify every detail about every variations in and new kinds of objects, properties, and events in the environment. Another important feature of these languages and ontologies is extensibility, allowing developers of systems in new areas to fill the ontological gaps, and then share their solutions with the community. Clearly, one should add, such languages will be of great value also in developing pervasive non-autonomous systems, such as IoT.

2. **Multi-Goal Management.** As mentioned earlier, autonomous systems have to deal with multiple goals. However, beyond the difficulty of specifying each goal separately, it is still extremely difficult to specify how the system should balance, prioritize, or weigh the competing goals under a bewildering multitude of circumstances. Even in a single given situation, and even if we allow the use of natural language, it is often close to impossible to state what the system should or should not do. Recalling, for example, the FFAV facing complex decisions with great human and business risks, we doubt that stakeholders can prescribe in advance what the system should do. In addition to these technical issues of specification, there are also weighty ethical issues. Courts of law often debate extensively whether a specific decision made by a human, in one particular situation, was right or wrong, negligent or not, in line with what is expected of a reasonable person. This becomes significantly harder in autonomous systems, which, for example, have to determine where to cut a tissue during a surgical procedure, or decide in a split second between two very bad alternatives in an emergency driving situation.

   Dealing with such challenges could be attempted first by adding (as expected) weights, priorities, and mutual constraints to individual goals and sets thereof, as well as endowing the system components responsible for the various goals with dynamic internal negotiation capabilities. A key role will also be played by mechanisms for specifying and handling contingent behavior, which reacts to and handles negative conditions directly related to the system's own actual behavior and choices. The ability to provide concise explanations of the system's decisions, both in real time and after the fact, will be of great value, allowing developers, and the system itself, to judge the programmed decisions and adjust them as needed.

   There is a very important issue that will have to be dealt with in a serious and deep way, in preparation for next generation autonomous systems, and which is clearly outside the scope of this paper. Since many kinds of such systems will, in a way, "make their own rules", the entire development process must take into account all relevant ethical/moral, legal, social, and political issues. Further, we will have to consider changes to systems (autonomous and non-autonomous) other than the one we are developing, and perhaps also to the rules and regulations of normal human life. Pedestrian traffic lights, and now-largely-illegal cellphone jamming in theatres, are examples of how humans may be externally forced to change their ways, in the presence of machines.

.
3. **<u>Capturing Emergent Behavior.</u>** During simulations, and often, subconsciously, stakeholders look for emergent properties and unexpected behaviors, which were not mentioned in the development process, and which may reflect problems or opportunities. For example, when observing the behavior of an autonomous vehicle, such as an FFAV or a golf cart, one may notice that sometimes it repeats a path with unusual precision, which creates unexpected wear on the floor, ground or grass. New requirements may be specified as a result, such as randomizing paths, documenting a limited set of supported surfaces, or taking advantage of this kind of predictability in other parts of the system. Also, humans are good at observing obstacles and near-accident conditions that the system should have noticed, but probably did not, again leading to improved specifications.

    Resolving this is not easy. It is not even clear how to specify a formal version of a trouble ticket, which describes an event, property or pattern that was noticed by the human but was not part of the original specification. Furthermore, in the STV&V stages (simulation, testing, verification and validation – discussed later), where, due to the enormous range of execution paths, extensive automation is expected, one would want to automate the detection of emergent properties as well. For anticipated behavior (desired or undesired) this would be quite similar to testing, but for unexpected behavior, automating the capturing itself in a formal, yet succinct and intuitive way, would appear to be impossible unless we have an adequate specification of unexpected behavior.

    In fact, we believe that Knuth's famous quote, "Beware of bugs in the above code; I have only proved it correct, not tried it," goes beyond recognizing the importance of testing given the limitations of formal methods and correctness proofs. It can be used to support our belief that testing a program is a must also because it yields totally new insights about what the program (and in our case, the autonomous system) does and does not do.

    Related to the issue of detecting and formalizing emergent behavior is, again, the issue of explainability and interpretability. This is especially relevant in the case of neural nets and of other 'black-box' solutions. Explanations summarize aspects of observed behaviors or of execution paths that dictate certain behavioral rules on certain inputs. In a way, explanations induce a model on the seemingly model-less machine-learning solution. And here too, summarizing such execution patterns automatically, in a way that is both formal and intuitive (as opposed to, say, presenting a large finite-state machine) is a challenging problem that is the object of active research.

To address these difficulties the foundation should thus offer:

1. <u>Domain-specific ontologies</u> for describing overall goals, desired and undesired behaviors and outcomes, and actual observed system and environment behavior.
2. <u>Languages,</u> within which these ontologies can be used with proper abstractions and generalizations, and which also allow for specifying weights, priorities, constraints, and other balancing techniques for considering multiple goals.

3. <u>Compilers and interpreters</u> for making these goal specifications executable, either as they are or after (automatically) decomposing them into agent/component-level goals.
4. <u>Emergent behavior monitors and detectors</u>, i.e., tools for automated extraction from sequences of events (or from videos or other sensor information) symbolic and concise descriptions of actual actions and behaviors (and patterns thereof) which are not mentioned in the specification and testing procedures as required, forbidden or allowed.

## 3.2    Controlling Execution in Simulation, Testing and Verification

Simulation, testing, formal verification of system models, and system validation against users' and customers' tacit needs and expectations (STV&V), will be of paramount importance for next generation autonomous systems. As is well-known, none of these techniques provides complete assurances, so they will have to be used in ways that complement each other.

Simulation provides flexible controllability and observability. It also facilitates observing emergent, both desired and undesired, not-yet-specified behaviors, in a variety of conditions. However, the simulated environment will always be an abstraction and simplification of reality.

Validating that a system does 'what the user wanted in the first place' is in itself a difficult task, since the real original wishes (and hopes) may remain tacit, and may or may not be reflected in the requirements documents and the specifications. The growing popularity of agile development methods stems, in part, from the fact that there are needs that users may not be able to state explicitly until they see an actual implementation or prototype that does not meet those very needs.

STV&V involves executing a system or a model thereof in a controlled manner, and/or traversing its possible states. It is an indispensable element of development, still, despite the availability of many excellent tools and methodologies, carrying out satisfactory STV&V for next generation autonomous systems calls for foundational work. The challenges stem from essentially two major issues:

**<u>The vast number of objects and variables:</u>**  Autonomous systems have to deal with a multitude of new objects and variables that are often ignored, simplified, or controlled in other systems. Each of these variables may itself be hard to control over the range of possible values. Just think of an FFAV deployed at a busy outdoors factory yard, with people, equipment, and vehicles moving around, with distinct shapes, colors, reflections, textures, sizes, locations, positioning and routes. Faithful testing or simulation of these system and environment is a formidable task.

**<u>The even greater number and intricacy of interactions</u>**: What the system itself does intentionally and explicitly may be reasonably controlled (e.g., in the case of the FFAV: pick up, travel, drop off, etc.). However, the number and complexity of the possible interactions and indirect effects between all objects in the environment and the system, are mind-boggling. Consider testing a very small FFAV working its way through a moving crowd of humans and machines, and perhaps other living creatures – as in a livestock show. The interactions are not constrained to "merely" reaching a destination, avoiding collisions and abiding by traffic laws (and between these there are already countless configurations and parameter values). Testing and simulation should be able to check and detect other kinds of interactions too. For example, whether the FFAV might splash passersby when it crosses a puddle, or if it can startle them – quietly and suddenly showing up at their side – or if it may get entangled in loose fabric or rope, or perhaps if it repeats certain route segments with unnecessary precision that causes the grass to wear out. Yet another relevant form of interaction is events causing the dynamic creation/appearance or destruction/disappearance of objects in particular circumstances, along with these objects' subsequent interaction with others.

The foundation should address the following key issues regarding STV&V,

### 1) Modeling Environments

Modeling environments should support the basic concepts that help specify, and then program, all these execution controls, in a consistent manner, across multiple application domains. They should integrate domain-specific libraries per various kinds of systems, tasks, environments, and contexts. Such libraries should be able to deal with the physical 3D space of real-world objects, basic motion and mobility, etc., in a variety of types of environments, such as transportation, medicine, home maintenance, manufacturing, person-to-person interaction, and many more.

Furthermore, modeling environments should offer adequate languages and their supporting tools for building system models from predefined heterogeneous components of agents and objects (at different levels of abstraction).

The modeling language should allow expression of coordination between components, including interaction between components as well as system reconfiguration features; e.g. creation/deletion of components, mobility and migration. The coordination language should be applicable to describing pre-known behaviors (e.g., knowledge and assumptions about the environment), as well as intended behaviors (e.g., those hopefully resulting from the programmed system).

The modeling language should allow achieving a desired level of realism by changing abstraction levels and simulation granularity. This implies multilevel and multiscale modeling, supporting heterogeneous and dynamically rich behavior – in particular, cyber physical behavior – and to do so on different levels of timing abstractions, from synchronous or timed to fully asynchronous or untimed.

Modeling languages should exhibit flexibility that goes far beyond that of traditional object-centric approaches, allowing, say, the description of group behavior (e.g., of a queue or a platoon), and focusing on the emergent interactions (e.g., collision or collision avoidance), rather than on single agents and objects and their individual perspectives (e.g., being at particular location or moving in a certain direction and at a certain speed).

Other abilities that the languages should support include specification of scenarios for controlling the execution as well as means for expressing clearly the simplifications made in the given specification compared to reality (for example, limiting variable ranges and environment actions, assumptions on the probability of certain events, etc.).

### 2) State-aware infrastructure for STV&V

Controlled execution and traversal of states following the above discussion requires a major infrastructure in its own right. The foundation we call for should provide this, via rich environments for common domains, and tools for enriching such setups, so that one can test and simulate autonomous agents in interaction with the complex cyber physical environment for which they are being built. The test and simulation environments should be 'state-aware' and convenient; i.e. they should communicate with engineers using HCI interfaces and logs that specify and describe states, conditions, actions, events, etc., in intuitive terms. This should include the state of the external environment, the internal state of the system and its agents, and the state of agents' perception of the external environment. For example, assume you want to figure out what the FFAV will do when it faces an obstacle consisting of two posts placed at a distance apart that is just a small amount above FFAV's width. Will it move between them or bypass them? Standard testing and simulation techniques call for actually placing the obstacles and observing the system's behavior. However, unless we also carefully check the feedback from the tested system as to what it 'thinks' it saw, one cannot be sure if it perceived the conditions as intended. For example, the FFAV may indeed pass between the posts, but for the wrong reason: it might have misclassified one or both of the posts.

Such perception control is, in fact, a particular case of a broader kind of state awareness, where the STV&V execution environment can report the system's own state and that of the environment; can guide the operation of certain system components based on the states of others; and can report, and react to, meta information about what the system does and does not do, the system's execution paths, etc. The complexity of this is amplified by the unpredictability of behavior and configuration. We believe that no explicit specification can cover the diversity and chaos of the real world. As a (relatively) simple example, how would we determine what states and interactions will occur in parallel to which others? Hence, the challenge remains: how does one create a simulator that can provide this kind of coverage?

**3) Testing**

Testing of simulated models or of real systems is of course critical. However, for autonomous systems that will be deployed worldwide in multitudes, we cannot expect testing to cover all possible circumstances. And as in formal verification, specifying and then checking that the response of the system-under-test to the given stimuli is indeed the expected one, covering all aspects of what was and was not done, and how it was done, presents a major challenge. We need metrics and criteria for evaluating the testing process. Software-testing criteria, such as coverage of the code (structural testing), meeting the pre-determined purposes of the testing (functional testing), and "volume of stimuli" (a sort of quantity-based testing – e.g., the number of miles driven in an AV test), are insufficient metrics for next generation autonomous systems. Instead, one of the things we would require, as a bare minimum, is a practical approach to measuring state coverage (for both system and environment), but here too, just the number or fraction of states covered will not suffice.

To improve testing efficiency we need techniques for automatic generation of sets of scenarios, subject to criteria that can be external, i.e. from the environment and the real world, or internal, such as coverage of specification and implementation entities. Furthermore, we need support for accelerated metamorphic testing in physical environments, which, roughly, means checking thoroughly that the system behaves correctly for a given scenario, and can then quickly providing assurances for many other scenarios that differ from the basic scenario only by small physical changes.

**4) Verification**

Formal verification may help guarantee desired behavior under all possible conditions, but due to the state explosion problem it may apply only to components or to simplified models of the entire system. Also, as described earlier, it is hard to fully formalize all aspects of the system under verification, and it is even harder to specify the assertions that describe the desired behavior in terms that are readily aligned with the expectations of the human users and engineers.

Verification and explanation of neural nets will be of paramount importance. We expect many next generation autonomous systems to have components based on machine learning and neural nets. To enable STV&V of such components, especially within complex systems, it is necessary to develop means for the formal verification of the correctness of neural nets, and for supplying adequate explanations of their internal behavior in terms aligned with the way engineers describe systems. These problems are long recognized as being very difficult, but there is an emerging field of research around them, whose initial results look very promising. To become part of our proposed foundation, however, the resulting techniques must reach a certain level of maturity and achieve broad acceptance.

## 3.3   Combining Model-Driven and Data-Driven Approaches

Conventional procedural software prescribes step-by-step processes and rules that are meticulously handcrafted and organized by humans. By contrast, machine learning (ML), and especially the use of neural nets (NN), results in complex machinery that is created by algorithms. There is now a growing call to find ways to combine the two techniques for modern systems, leveraging their relative advantages to complement each other [10, 22]. There is still a severe lack of detailed guidelines on how to do this, the combination being very different from integration practices in classical engineering. Due to the new challenges involved (see Section 2.3), the problem is exacerbated for next generation autonomous systems.

Below are some of the differences between developing the traditional procedural, model-based solutions and the construction of ML-based ones, which must be bridged in order for one to be able to integrate the results of the two in a single cohesive system. To better concentrate on the integration issue, we make the assumption that many of the specific still-open research problems in each of them (like verification and explainability of neural nets) are already resolved.

1) **General life cycle.** The stages that traditional software engineering methods call for – requirements elicitation and specification, design, code, testing, etc., whether in the traditional waterfall and V-shaped models or agile/spiral ones – do not yet include activities related to machine learning, such as the collection, validation and sampling of training data, the actual training (with evaluation and re-training), etc.

2) **Requirements specification.** There are deep differences between requirements specification in classical software engineering and in the ML setting, and these issues are distinct from the general issues discussed in Section 3.1 around requirements for next generation autonomous systems3.1. Consider even simple specifications, like the one stating that an electrical switch must close when the temperature reaches 80 degrees, or the one stating that the brakes must be activated when a stationary obstacle is sensed and the stopping distance at the current speed is less than 1 second. These are well defined, and engineers can translate them easily into working components in more than one way. However, for a system containing an end-to-end component that was trained to handle excessive heat or avoid collision based on positive and negative examples, it is not at all clear how to specify what the component has to do, either before it is built or when it tested. This issue is also related to the explainability and interpretability of ML, discussed next.

3) **Explainability and interpretability.** Generally speaking, explainability and interpretability of a system refer to the ability to describe *what* the system does, articulate the underlying *mechanisms* it uses (i.e., algorithms, computations, etc.), and/or *justify a given decision* it takes. Neural nets can be very successful for complex tasks, but their internal workings are often a mystery, resulting, as they were, from ML techniques that use extensive training and adjustments. While much research is being devoted these days to explainability and interpretability of neural nets, it is still a far cry from the situation with traditional programming, for which engineering practices recommend writing code that is easy to read and understand, and to enhance it with ample comments. We believe that for very large neural nets, even when explainability/interpretability tools are eventually able to extract the tacit rules behind the arcane computation, the relation between these rules and the neural net's actual mechanisms will be very different from the relation between the source code of a conventional module in a high level programming language and lower level code into which it was compiled.

4) **Decomposability.** Closely related to the differences around explainability are differences around the mindsets associated with the traditional model-based and the ML data-based approaches. In model-based designs, most system artifacts can be hierarchically decomposed into well understood functional and structural elements, the role of which in the local function or in the bigger picture is more-or-less clear. Such decomposition is desired during almost all stages of the development effort. On the other hand, designing ML data-based solutions is typically accompanied with an end-to-end mindset, be it for an entire system or for a particular problem. Decomposing a machine-learning solution into meaningful parts may be an interesting challenge derived from explainability or verification goals, but it is not an intuitively expected result of the actual development of the solution.

5) **Testing and verification.** The difficulty in specifying the behavior of neural nets makes it very difficult to test or to verify their correct behavior. While some exciting work has been done on checking properties thereof (see, e.g., [13] , the very concepts of testing and verifying neural nets still require extensive research.

6) **Trustworthiness and Certification**. Trustworthiness and certification are closely related, of course, to testing and verification, but they deserve a separate discussion. Many kinds of autonomous systems are highly critical---Failure to meet their expected behavior can be disastrous. Currently, critical system design calls for taking special care to provide the trustworthiness guarantees that are required by standards, e.g. DO178B, FAA certification and ISO 26262 for the safety of electronic components in the automotive industry. The engineering of most critical systems relies on model-based techniques that give rise to predictability at design time. However, components based on machine learning are not engineered in the same way, and achieving an accepted level of trustworthiness and certification requires new technical solutions. The complexity and unpredictability associated with next generation autonomous systems prevent present end-to-end testing and simulation techniques from being adequate. Complementary non-technical measures, such as risk management, the concept of insurance, or the use of the justice system as deterrent against negligence, are a separate issue altogether and are outside of the scope of this paper.

Consider a proposed solution that is to ultimately consist of both model-driven components (based, e.g., on an object model, algorithms, scenarios, rules, and decision tables) and a ML data-driven components (based, e.g., on neural nets), where each addresses a separate part of the problem space. At some point, the engineers have to decide which sub-problems should be solved using which of the two approaches. Sometimes the answer is easy: for example, reading traffic signs may be fully ML-based, while the decision to remain below a known speed limit can be model based. Staying in lane on a clean well-marked highway might be model-based, while negotiating a road surface covered with sand should probably be ML-based. However, in many cases the choice between these two approaches will be a lot more difficult. Besides, the means for deciding which situation is the relevant one in order to activate the proper component (e.g., whether the road surface is clean or sandy), will itself be either via model-based or ML-based, and this has to be decided upon by the designers too.

## 4   Related Work

The existence of gaps between state of the art and achieving desired trustworthiness has been articulated, e.g., by Neumann [20] and Bellovin and Neumann [2], with a focus on component vulnerability, the impact on composite-system vulnerability. In these papers the authors propose that certain classical, general engineering principles, be applied to help address these. The current paper expands on this discussion of general principles, and proposes building concrete new elements of an engineering foundation, and of models of autonomous systems and their environments, which could be directly used in enhancing systems trustworthiness.

Many definitions have been offered for autonomy and agency (see, e.g.,[16, 9]). In Chapter 2 we have introduced definitions of autonomous systems concepts to enable the ensuing discussion. Autonomous systems have been studied in many contexts and perspectives. *Autonomic Computing* (see, e.g., [5]) focusses on computer systems and networks capable of self-management, and in particular, automating dynamic configuration. The broad research of *multi-agent systems* (see, e.g., [23]) pays a special attention to how (autonomous) agents can combine local goals with collaboration rules and distributed algorithms to achieve system-wide overall goals. Autonomy is often associated with *self-awareness* (see, e.g, [14]), which then translates to a system being able to use its own overall states as input to further decisions. *Symbiotic Computing* (see, e.g., [6]) studies how autonomous systems can interface collaborate with humans and complex organizations, discussing many technical, commercial, and ethical implications. Further, entire conferences and journals are dedicated to the topic of autonomous systems (see, e.g., [1]). However, in these bodies of work about how to build and test complex, autonomous, safety-critical systems, it is still hard to find theories and tools for building confidence (or trust) in the system's totally-unsupervised decision making when it faces unpredictable conditions and events.

In testing autonomous systems the focus is achieving coverage: of behavioral elements of the system itself (e.g., all its potential actions and sequences thereof), of physical system components (methods and lines of code), and of the already known and already predicted conditions and events in the environment. E.g., DeepTest [27], is able to discover undesired behaviors in deep neural nets by analyzing deep neural nets and intelligently exploring the effects of changes in the known conditions. Another angle of the effort to reach extensive coverage, can be seen in the how makers of autonomous vehicles present the number of real and simulated miles driven by these vehicles. This should imply that random and real world scenarios that the tested system experienced would exceed dramatically those experienced by a driving student or by a typical driver throughout their lifetime (see, e.g., [15],[28]). However, additional information is needed in order to independently confirm that the probability that any autonomous vehicle around the world will encounter an untested scenario, and will mishandle it, is sufficiently small.

We are not aware of extensive and systematic engineering handling of the very principles that stand behind unpredictability. Nevertheless, demonstrations and discussions of autonomous systems, with their failures and successes, may implicitly highlight not just how complex, but how unpredictable, a situation may be. E.g., in the SpotMini video [3] (and in the critical discussion of its demonstration environment in [21]), one can see how a soda can that is accidentally opened causes confusion, and be alerted to the myriad other things that could happen.

Directions towards common domain-specific languages, ontologies, models and behavioral scenarios in the area of autonomous systems can be seen in, e.g., Traffic Sequence Charts language [7], USA National Highway Traffic Safety Administration (NHTSA) scenarios [18], Autoware open source software for autonomous vehicles [12], and open source simulators like CARLA [8]. These solutions however are not integrated with each other, each using its own terms and concepts, as building blocks in a bottom-up tool construction. We intend for the proposed foundation to be provide a top down approach for concepts that cut across all aspects of autonomous systems and autonomous behavior.

There are numerous simulation tools for autonomous systems, see, e.g., for autonomous driving, open source CARLA [8], Paracosm [17], HEXAGON MSC [11], and COGNATA [4]. Such simulators can provide synthetic and/or real-world sensor data (like camera images, GPS and LiDAR) to a driving ego car, which feeds back its actions like acceleration and steering, which in turn allows the simulator to update the scene accordingly. Some of the tools also offer a physics engine – which automatically enhances behaviors. Simulators offer a variety of built-in scenes and scenarios, as well as APIs/languages and interactive tools for engineers to further control the test environment. The PARACOSM tool provides an additional key capability: systematically exploring the environment-parameters' space (including combinations therein), through planned sampling, in search for undesired behavior. While such simulators are important in testing and otherwise observing autonomous behavior, the proposed foundation calls for enhanced and additional important capabilities, like (a) substantial extension of the range of user control over environment variability as presently synthetic scenes often appear over-simplified, and scenes derived from real images are less controllable; (b) the ability to externally create a mapping of how the simulated environment is reflected in the 'mind' of the system under test (=SUT; the ego car in the case of AV); (b) automated detection and evaluation of emergent properties of the behavior of the SUT, and which was not in the test specification (e.g., was the driving of the tested car erratic in some not-yet-specified way? or, were other road users unreasonably forced to change their behavior in order to accommodate the tested car behavior?). Another limitation is the fact that many sophisticated functions are presently spread across multiple tools. Approaches for solutions are described in [19], in the context of the CPS Wind Tunnel tool (CPSWT), the MOSAIK smart grid simulator, and Modelica's Functional Mockup Unit (FMU).

The proposed foundation will both draw from, and contribute to, the many published results, working tools, and industry standards, that already deal with various perspectives of autonomous systems. In its more straightforward (but still significant and hard to accomplish) aspect, the foundation will consolidate best practices, distill underlying principles, and highlight unique solutions. However, its greater contribution will be in that it will introduce new solutions for topics like: how to specify (or document) desired and observed properties and behavior in an ever-changing (and unpredictable) environment; how to create complex, yet transparent, domain-specific models, with abstractions that accommodate entities that do not yet exist; and, how to be able to express, react to, and control, the state of the system and its environment in requirements, code, tests, simulations and documents.

In this paper we have argued that closing certain gaps is mandatory step on the road to development and ubiquitous deployment of next-generation autonomous systems. We are well aware that there could be a wide range of opinions by engineers, company leaders, governments and regulators, and the public at large, about what measures are absolutely necessary to achieve basic trustworthiness. Combinations of insurance to compensate for errors and omissions, and reliance on a justice system to strengthen stakeholder accountability, may indeed enable deployments of autonomous systems using current practices without technological breakthroughs. Testimony to such risk-related deliberations appears, e.g., in an article covering the Waymo-vs-Uber case [29] (which we cite here without expressing our own opinion on the case or on the quoted positions). Another alternative measure is of course constraining system autonomy, and keeping ultimate human responsibility as an essential part of decision making processes. In this paper we argue that the new foundation should deal directly with concrete measures for dealing with trustworthiness, in the face of complexity and unpredictability. This should, in turn, accelerate the deployment and acceptance of systems that can be considered autonomous.

# 5   CONCLUSION

Next-generation autonomous systems are imminent, given the obvious need to further automate existing organizations, by gradually replacing human operators by autonomous agents. Such systems are often critical and are expected to manage dynamically changing and possibly conflicting goals, which reflects the transition from "narrow" or "weak" AI to "strong" or "general" AI. These systems will have to cope with the uncertainty of complex, unpredictable cyber physical environments, and will have to adapt to multiple goals. And they will have to harmoniously collaborate with human agents, giving rise to so-called "symbiotic" autonomy.

Classical software and systems engineering will need to be thoroughly revised. Existing design techniques are model-based, mostly applicable to small sized and centralized automated systems, with predictable environments and well-understood specifications. Such techniques cannot be applied as is to next generation autonomous systems, for many reasons: They inevitably rely on the use of ML-enabled components, which, as a rule, do not admit model-based characterizations, and their geographic distribution and mobility require network infrastructure of extremely high trustworthiness, which is a far cry from the current situation.

Autonomous vehicles provide an emblematic topical case, illustrating the challenge that involves huge economic stakes and deep societal impact. Public authorities allow "self-certification" for autonomous vehicles, whose manufacturers often adopt "black-box" ML-enabled end-to-end design approaches. They frequently consider statistical evidence to be enough for trustworthiness: "I've driven a hundred million miles without an accident, so it's safe." Furthermore, critical software can be customized by updates which may be designed and added with even less scrutiny. The software for Tesla cars, for example, may be updated on a monthly basis.

All this has generated lively public debates. Many important voices tend to minimize the risks from the lack of rigorous design methods: Some claim that we should accept the risks because the benefits will be far bigger! Others exhibit blind faith in empirical methods and argue that rigorous approaches to complex problems are inherently inadequate. Some people are overoptimistic, arguing that we really do have the right tools, and it is just a matter of time.[3]

Our central argument is that advent of next generation autonomous systems raises an extraordinary scientific and technical challenge. The degree of success in meeting this challenge will ultimately determine the extent of acceptance of such systems, as a compromise between their estimated trustworthiness, the anticipated benefits of the automation they afford, and the required changes in other systems and in human behavior.

---

[3] "I almost view [autonomous cars] as a solved problem. We know what to do, and we'll be there in a few years." E. Musk, Nvidia Technology Conference, March 2015.

The paper's contribution is twofold. First, we propose a basic terminology for next generation autonomous systems, and a framework capturing their main characteristics. The framework provides insight into the spectrum of possibilities between automation and autonomy, and is intended to help in understanding the degree of autonomy of a system as the division of work between a system and human agents.

Second, we advocate the need for a new scientific and engineering foundation that will address the key open issues in the engineering of next generation autonomous systems. Such an autonomics foundation will hopefully lead to trustworthy hardware/software systems, which are reactive and integrate model-based cyber physical and ML-enabled components. We argue that such a foundation cannot be obtained merely by combining these approaches, which have focused on software systems that, for the most part, fall under the headings of autonomic computing, adaptive systems and autonomous agents.

We anticipate that forming this foundation will require major and ground-breaking efforts in three main directions.

The first direction is to develop a rigorous theory and supporting tools for dealing with heterogeneous specifications. These should make it possible to characterize system behavior in a broad fashion, including the behavior of its individual agents as well as the system's global behavior in terms of its overall goals and its emergent properties.

The second direction is aimed at providing sufficient evidence of a system's trustworthiness. Existing formal verification methods are vastly inadequate, being applicable only to subsystems of moderate complexity and suffering from well-known limitations aggravated by the use of ML-components. Hence, we emphasize the paramount importance of modeling and simulation: on the one hand, faithful, realistic modeling is needed so that behavior will be accurate and will appear true to life; on the other hand, semantic awareness is needed so that the experimenter has access to a meaningful abstraction of the system's dynamics, allowing controllability and repeatability of the testing. The latter will also allow notions of coverage that measure the degree to which relevant system configurations have been explored.

The third direction of the effort required for the foundation involves moving from traditional system design to "hybrid" design, seeking tradeoffs between the trustworthiness of classical model-based approaches and the performance of data-based ML ones. Currently, the less-than-adequate applicability of rigorous model-based approaches to complex systems has led to the development of end-to-end ML-enabled autonomous systems, which comes too with major disadvantages. Taking better advantage of each approach requires the development of common architectural frameworks that would integrate modules characterized by their pure functionality independent of their design approach. Developing the theory for decomposability and interoperability as well as explainability of data-based modules is essential for reaching this goal.

In summary, we are definitely at the beginning of a revolution, where machines are called upon to progressively replace humans in their capacity for situation awareness and adaptive decision making. This is a first step toward general AI, which goes far beyond the objectives of ML-enabled intelligence. The role of autonomous systems will depend on choices we make about when we trust them and when we do not. Giving ourselves the means to make informed decisions is essential.